\documentclass[]{spie}  
\usepackage[]{graphicx}

\title{Optical Spectroscopy of the Hole Spin in GaMnAs Acceptors} 

\author{B. Lakshmi, G. Favrot and D. Heiman
\skiplinehalf
Department of Physics, Northeastern University, Boston, MA 02115}

\authorinfo{Further author information: (Send correspondence to d.heiman@neu.edu)}
 
\begin{document}  
\maketitle 

\begin{abstract}
The spin state of holes bound to Mn acceptors in GaMnAs is investigated by optical spectroscopy. 
Concentrations of Mn from 10$^{17}$ to 10$^{19}$ cm$^{-3}$ were studied as a function
of magnetic field and temperature.  The photoluminescence from recombination of electrons with
holes bound in the Mn-acceptor complex (MAC) displays multiple spectral peaks. The circular
polarization of these peaks increases with increasing magnetic field and saturates at
$\rho\simeq$1/3. This value of polarization is expected from modeling the addition of spin angular
momentum and interband optical transition matrix elements.
\end{abstract}


\section{INTRODUCTION}

Although conventional electronics rely on the transport and storage of electron \textit{charge},
new devices for information technology are expected to rely on the transport and storage of electron
\textit{spin}.  The emerging field of "spintronics" could spawn devices such as spin transistors or
spin valves \cite{Datta1990}, spin memory devices,\cite{Prinz1998} and even quantum spin
computers \cite{Loss1998,Kane1997}.  Of particular interest are new materials which have
magnetic properties as well as semiconducting properties. Electronic devices based on GaAs are
becoming increasingly useful in high speed electronics, most notably in the area of wireless
communication where frequencies are above a GHz.  GaAs doped with $\sim$5\% Mn is
ferromagnetic with a Curie temperature in excess of 100 K. \cite{Ohno1996,Ohno1998} This
III-V based magnetic semiconductor is grown by low-temperature molecular beam epitaxy
(MBE).  The ferromagnetism is generated by a high concentration of holes which are contributed
by Mn acceptors.  Theories of ferromagnetism in this material have been based on RKKY
\cite{Dietl2000,Konig2000} or double exchange \cite{Akai1998} mechanisms. The emergence of
ferromagnetic GaMnAs has spawned renewed interest in magnetic semiconductors.

At the heart of the ferromagnetism in GaMnAs is the Mn acceptor and the interaction between the
spins of acceptor-contributed holes and spins of Mn$^{2+}$ ions.  Table~1 lists the Mn states in
GaAs and GaP semiconductors.  At low concentrations the neutral acceptor (A$^0$) exists ---
substitution of Mn$^{2+}$ for Ga$^{3+}$ generates a negatively-charged Mn$^{2+}$ core
which binds a valence band hole. At high Mn concentrations the (A$^0$) is not observed since the
bound-hole wavefunctions overlap and become unbound.\cite{Szczytko1999c} The ionized acceptor 
(A$^-$) is found in GaAs samples at both low and high Mn concentrations.

\begin{table} [h]
\begin{center}
\caption{Mn impurities in GaAs and GaP semiconductors.}

\vspace{0.1cm}
\begin{tabular}{ccccc}
\hline
\hline
 Impurity & Ion & Configuration & Material & Ref. \\
\hline
A$^0$ & Mn$^{2+}$ & ($d^4$+$e$)+$h$ & GaAs bulk   &
\cite{Schneider1987,Szczytko1999c} \\
A$^-$ & Mn$^{2+}$ & ($d^4$+$e$)           & GaAs bulk     &
\cite{Schneider1987,Szczytko1999c} \\
  ''  & ''        & ''                    & GaAs epilayer & \cite{Szczytko1999c} \\
 --  & Mn$^{3+}$ & $d^{4}$                 & GaP           & \cite{Kreisel1996} \\
\hline
\hline
\end{tabular}
\end{center}
\end{table}

\subsection{Magnetic Binding Energy of Acceptor}
The hole bound to Mn has a large binding energy (113~meV),\cite{Chapman1967,Petrou1985}
several times larger than for an effective mass acceptor such as carbon (26~meV). The larger binding 
is attributed to so-called "central cell" potential. The central cell binding is mainly from 
additional \textit{chemical} binding. However, there is also \textit{magnetic} binding of the hole 
to the Mn due to the \textit{p-d} exchange interaction, amounting to 26~meV.\cite{Bhattacharjee1999}
The 4-times larger ground state binding energy is correlated with a smaller effective Bohr radius 
for the hole bound to the Mn ion.\cite{Bhattacharjee1999} In a simple hydrogenic model, the Bohr 
radius for A$^{\rm EM}$ is \textit{a}$_{\rm EM}$=22~\AA , as obtained from \textit{a}=(0.53~\AA ) 
$\epsilon$/\textit{m}$_{{\rm h}}$, using the dielectric constant $\epsilon$=12.6, and 
\textit{m}$_{\rm h}$=0.3 from E=(13,600~meV)\textit{m}$_{\rm h}$/$\epsilon$$^{2}$. For A$^{\rm Mn}$, 
the characteristic radius of the wavefunction for lightly doped bulk-grown material is expected to 
be only \textit{a}$_{\rm Mn}$=10~\AA , obtained from the metal-nonmetal transition at a hole 
concentration of p$_{\rm c}$$\sim$ 2$\times$10$^{19}$ cm$^{-3}$ and the Mott criteria 
\textit{a}p$_{\rm c}^{1/3}$=0.25.\cite{Woodbury1977} However, in MBE grown Ga$_{1-x}$Mn$_x$As 
\cite{Ilegems1975} a metal-insulator transition appears at x=0.03,\cite{Iye1999} corresponding to a 
much higher hole concentration, p$_{\rm c}\sim{\rm xN}_{\rm o}\sim7$$\times$10$^{20}$~cm$^{-3}$ 
(N$_{\rm o}$=2.2$\times10^{22}$~cm$^{-3}$), assuming one hole per Mn ion. This would give a Mott 
radius of only 3~\AA, even smaller than the 4.0~\AA~nearest-neighbor Mn-Mn cation spacing.


\subsection{Magnetic Acceptor Complex}

The \emph{magnetic acceptor complex} (MAC) is composed of the j=3/2 hole and S=5/2
Mn$^{2+}$ (3d$^{5}$) ion.\cite{Liu1995} The spin configuration of the MAC is dominated by
the \textit{p-d} exchange coupling. In the present work, magneto-optical experiments are used to
probe the spin configuration of the magnetic acceptor complex in lightly doped GaAs:Mn.  A
magnetic field is applied in order to align the total spin of the MAC
(\textbf{J}=\textbf{j}+\textbf{S}).  The spin configuration is measured via the polarization of
the recombination radiation as the bound hole recombines with photoexcited electrons. Both
circular polarization in a Faraday configuration and linear polarization in a Voigt configuration 
are studied. Analysis of the polarization as a function of magnetic field and temperature is used to
obtain the spin configuration of the MAC. These results are compared to modeling which takes
into account the angular momentum coupling as well as transition matrix elements of the
recombination.

\subsection{Acceptor Exchange Energy}

The \emph{p-d} exchange interaction between hole spins and Mn-ion spins has an energy of 
$\beta$N$_{0}$$\approx$~$-$2~eV. \cite{Szczytko1999b}  The minus sign reflects the antiferromagnetic 
exchange between the spin vectors.  Table~2 lists values for $\beta$N$_{0}$ obtained by various 
methods.  Values range from $\beta$N$_{0}$=~$-$1.2 to $-$3.3~eV.  The optical methods rely on 
combining observed splitting of optical transitions with magnetization 
data.\cite{Heiman1983,Heiman1997} The major source of error is normally due to the uncertainty in 
measuring the optical splitting. In GaMnAs the splitting is much smaller than the broadening of the 
transitions, making it difficult to accurately measure the splitting. \cite{Szczytko1999b} 
Magnetotransport measurements rely on fitting the magnetic field dependence of resistivity to a 
model of electronic scattering which contains the hole effective mass and carrier concentration. 
\cite{Matsukura1998c,Omiya1999}  There, the major source of error is due to the uncertainty in the 
assumed hole effective mass (\emph{m}$^{*}_{h}$/\emph{m}$_{o}$=0.3-0.5) and the hole concentration. 
Hole concentrations can be obtained from Hall effect measurements, but complications arise in GaMnAs 
from additional contributions to the Hall resistance due to the extraordinary Hall effect and 
negative magnetoresistance. These effects become negligible at low temperatures (T=50~mK) and high 
magnetic fields (B$>$20~T).\cite{Omiya1999} However, it is not clear whether the resistivity 
accurately samples the entire population of holes which generate the ferromagnetism.

\begin{table} [h]
\begin{center}
\caption{\emph{p-d} exchange energy in GaMnAs.}
\vspace{0.1cm}
\begin{tabular}{cll}
\hline
\hline
$\beta$N$_{0}$ & Method & Ref.\\
\hline
$-$2.5~$\pm{0.2}$~eV & exciton splitting & \cite{Szczytko1999b}\\
$-$1.2~$\pm{0.2}$~eV & core-level photoemission & \cite{Okabayashi1998}\\
$\mid$3.3$\mid$~$\pm{0.9}$~eV & magnetotransport & \cite{Matsukura1998c}\\
$\mid$1.5$\mid$~$\pm{0.2}$~eV & magnetotransport & \cite{Omiya1999}\\
$-$2.1~$\pm{0.2}$~eV & optical absorption & \cite{Szczytko1999b}\\
\hline
\hline
\end{tabular}
\end{center}
\end{table}

Note that the exchange \emph{energy} $\beta$N$_{0}$ refers to the exchange interaction between 
\emph{itinerant} holes and Mn ions.  $\beta$N$_{0}$ is related to the to the \emph{local} exchange 
constant $J_{p-d}$ in the acceptor defined by E=$J_{p - d}(\textbf{S} \cdot \textbf{j})$.  It is 
difficult to determine the explicit relation between the two quantities because the holes are 
tightly bound to acceptors where confinement leads to higher wavevector components of the 
exchange.\cite{Bhattacharjee1999}
\section{EXPERIMENTAL BACKGROUND}

The GaAs:Mn samples contained a thin layer ($\sim$~$\mu$m) of Mn doping which was
produced by diffusion.  MnAs was deposited on the surface of semi-insulating GaAs then heated
to high temperatures (800-900~C) for several hours.  Details of the diffusion procedures have
been published elsewhere.\cite{Wu1992} Capacitance-voltage (C-V) electrochemical etch profiler
and secondary ion mass spectroscopy techniques were used to determine the Mn concentration
profile which usually contained several plateaus of order $\sim$~$\mu$m each.  In order to
achieve the desired concentration some of the sample surface was etched away.

Photoluminescence (PL) experiments were carried out using a low-power 0.63~$\mu$m He-Ne laser which 
was unfocused and provided a power density on the sample of 10$^{-1}$ W/cm$^{-2}$.  When the
laser wavelength was shortened to 0.5~$\mu$m in order to obtain a smaller penetration depth of the 
laser, no changes were observed in the spectra.  Measurements were made in a optical cryostat 
containing a B=9 T superconducting magnet solenoid.  The axis of the split-coil solenoid was 
horizontal, permitting light to be directed parallel or perpendicular to the field direction, for 
Faraday or Voigt geometries, respectively.  Circular polarization was selected by an infrared 
Polarcor dichroic glass linear polarizer and a $\lambda$/4 Polaroid retardation plate ($\lambda$/2 
at 0.4~$\mu$m).  This combination produced polarization rejection exceeding 95\% in the wavelength 
region of interest.  Opposite circular polarizations were obtained by rotating the retardation plate 
by 90~deg. After the polarization was selected, the PL light was focused into a 600 $\mu$m diameter 
core optical fiber which transferred the light to a 1/4~m imaging spectrometer and CCD camera.

Polarized spectra were obtained in Faraday and Voigt configurations, with light propagating
along or at right angles to the magnetic field direction, respectively.  Circular polarization in 
the
Faraday configuration is defined in the usual way,
\[\rho_{{\rm cir}} \equiv (I_{+} - I_{-}) / (I_{+} + I_{-}),\]
where $I_{+}$ and $I_{-}$ are the PL intensities for light polarized $\sigma_{+}$ and
$\sigma_{-}$, respectively. In a similar way, linear polarization in the Voigt configuration is
defined by,
\[\rho_{{\rm lin}} \equiv (I_{\|}-I_{\bot}) /
(I_{\|}+I_{\bot}),\] where $I_{\|}$ and $I_{\bot}$ are the PL intensities for light polarized
parallel and perpendicular to the magnetic field direction, respectively.

\section{EXPERIMENTAL RESULTS}
\subsection{Electron-hole Recombination Emission}
\begin{figure}
   \begin{center}
   \begin{tabular}{c}
     \includegraphics[height=8cm]{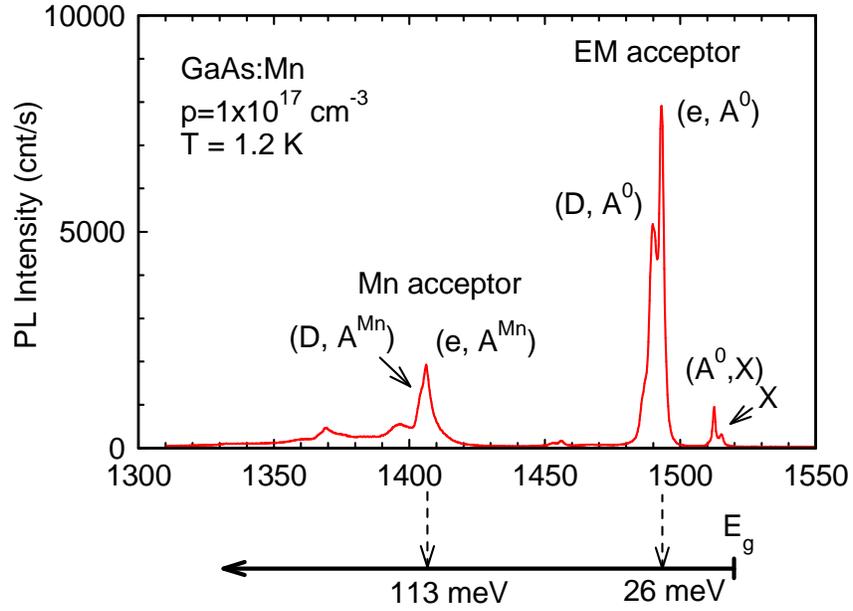}
   \end{tabular}
   \end{center}
\caption{Photoluminescence intensity versus photon energy for GaAs:Mn, with Mn concentration
1$\times$10$^{17}$ cm$^{-3}$ and T=1.2~K. The spectral peaks correspond to recombination
of electron-hole states described to in the text.}
\end{figure}

The PL spectrum for a lightly Mn-doped (1$\times$10$^{17}$~cm$^{-3}$) GaAs sample is
shown in Fig.~1. There are three spectral regions of interest.  These regions correspond to
recombination emission of valence band holes in various configurations.  Near the bandgap,
between 1510 and 1520 meV, holes are bound in \textit{excitons}. The dominant peak in this
region arises from recombination of excitons bound to effective mass acceptors, denoted
(A$^{\rm EM}$,X).  In the region between 1480 and 1500~meV, holes are bound to
\textit{effective mass acceptors}, so recombination arises from electrons which are either free,
(e,A$^{\rm EM}$), or bound in donors, (D$^{\rm o}$,A$^{\rm EM}$). Table~II displays the
transitions and their energies.

\begin{table}
\begin{center}
\caption{Transition energies of electron-hole recombinations in lightly Mn-doped GaAs:Mn.}
\vspace{0.1cm}
\begin{tabular}{ll}
\hline
\hline
 Energy (meV)& Transition\\
\hline
1515.17 & X (free exciton)\\
1512.5 & (A$^{\rm EM}$,X)\\
1493.0 & (e,A$^{\rm EM}$)\\
1489.7 & (D,A$^{\rm EM}$)\\
1456.0 & (e,A$^{\rm EM}$)-- LO\\
1452.9 & (D,A$^{\rm EM}$)-- LO\\
1406.2 & (e,A$^{\rm Mn}$), (D,A$^{\rm Mn}$)\\
1396.1 & A*$^{\rm Mn}$\\
1369.2 & (e,A$^{\rm Mn}$)-- LO\\
1359.8 & A*$^{\rm Mn}$-- LO)\\
\hline
\hline
\end{tabular}
\end{center}
\end{table}

In the spectral region of current interest, near 1400 meV, the recombining holes are tightly bound 
in acceptor states with Mn at the center.  Similar to the effective mass acceptors, electrons are 
either free (e,A$^{\rm Mn}$) or bound to donors (D$^{\rm o}$,A$^{\rm Mn}$). 
\cite{Bimberg1978,Petrou1985} The peaks labeled A*$^{\rm Mn}$ have the same
polarization as the other A$^{\rm Mn}$ transitions in the 1400~meV region and are attributed to
Mn acceptors with an additional 10~meV binding of unknown origin. There are also features 
corresponding to LO phonon replicas of the acceptor features which are downshifted by 36.7~meV. The 
scale at the bottom of Fig.~1 shows the spectral energies relative to the bandgap at 1519~meV. From 
these energies, the binding energy for the hole in Mn acceptors is seen to be 113~meV, while that 
for the effective mass acceptor is only 26~meV.

Figure~2 displays PL spectra for other Mn concentrations, from 10$^{17}$ to 10$^{19}$
cm$^{-3}$. Both effective mass and Mn acceptor features are observed.  The intensity of the
Mn-related features decreases with increasing Mn concentration. This is attributed to several
factors.  At the highest concentration, 10$^{19}$ cm$^{-3}$, the acceptor wavefunctions are
beginning to overlap, leaving fewer Mn acceptors which are isolated.  In addition, at higher 
concentrations there is greater recombination from effective mass acceptors, due to higher disorder.

\begin{figure}
     \begin{center}
     \begin{tabular}{c}
     \includegraphics[height=6cm]{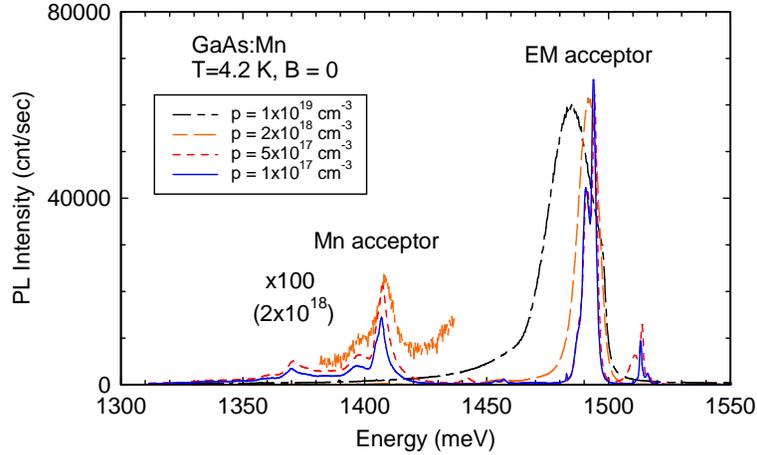}
     \end{tabular}
     \end{center}
     \caption{Photoluminescence spectra for GaAs:Mn with various Mn concentrations.}
\end{figure}

\begin{figure}[b]
     \begin{center}
     \begin{tabular}{c}
     \includegraphics[height=5.5cm]{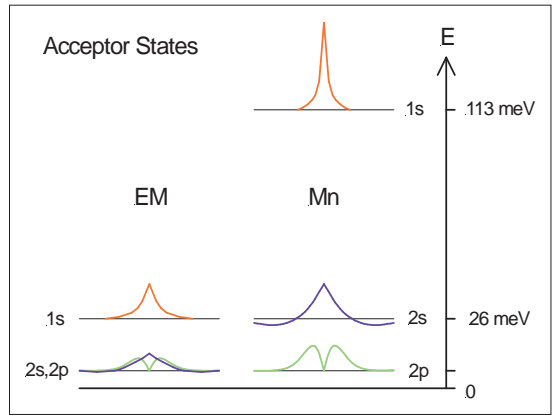}
     \end{tabular}
     \end{center}
     \caption{Diagram of effective mass (EM) and Mn acceptor states in GaAs:Mn.  Energy
levels and wavefunction profiles are shown for various hydrogenic acceptor states.}
\end{figure}

Figure~3 displays the energy levels relative to the valence band for both EM and Mn acceptors. On 
the left side, the 1S(EM) orbital level is at 26~meV.  On the right side, the 1S(Mn) level is at 
113~meV.  The larger binding energy for the Mn acceptor compared to the effective mass
acceptor results from additional "central cell" type attraction.  The 2S(Mn) level, at 25.3~meV, 
\cite{Linnarsson1997} is close to the 1S(EM) level.  In contrast, the 2P(Mn) level is far removed 
from the 2S(Mn) level, being close to the 2S/2P(EM) levels.  This large splitting between S- and 
P-states for the Mn acceptor has been observed in infrared excitation 
experiments.\cite{Linnarsson1997}  Both the \textit{coincidence} of the energies of the 2P(Mn) and 
2S/2P(EM) levels and the large 2S(Mn)-2P(Mn) \textit{splitting} result from the character of the 
P-orbital wavefunctions of the hole, illustrated in Fig.~3.  The wavefunction of the P-orbital is 
zero at the Mn ion site which inhibits the central cell attraction --- point-contact central cell 
attractive potentials are effective only when the hole is at the position of the Mn ion. This holds 
for both chemical and magnetic central cell attraction.

\subsection{Polarization of Emission}

The PL spectrum was measured in an applied magnetic field, B. Shifts in the spectral energy were
small, less than 1~meV for fields up to B=9~T.  The large shifts and splittings reported in a
previous study \cite{Liu1995} could not be reproduced here. In spite of the small energy shifts,
the spectra showed large changes in the \emph{polarization}.  Figure~4 displays spectra at
B=2~T in the two circular polarizations, $\sigma_{+}$ (solid curve) and $\sigma_{-}$ (dashed
curve).

\begin{figure}
\begin{center}
\begin{tabular}{c}
\includegraphics[height=8cm]{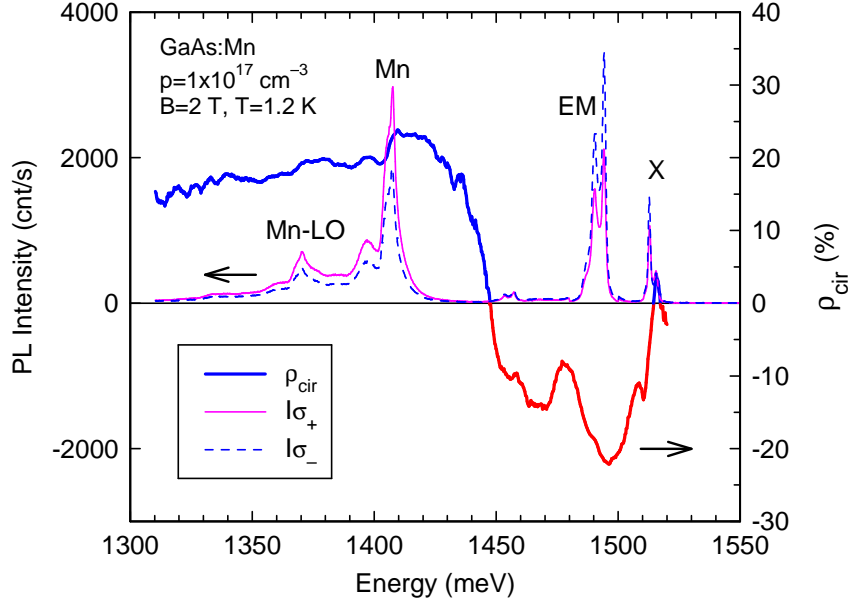}
\end{tabular}
\end{center}
\caption{Photoluminescence intensity versus photon energy for GaAs:Mn, with Mn concentration
1$\times$10$^{17}$ cm$^{-3}$, at B=2~T and T=1.2~K.  The thin solid (dashed) curve
corresponds to $\sigma_{+}$ ($\sigma_{-}$) circular polarizations. The thick solid curve
corresponds to the circular polarization $\rho_{\rm cir}$.}
\end{figure}

In the region of the effective mass acceptor, labeled EM, the emission is stronger in $\sigma_{-}$ 
polarization than in $\sigma_{+}$.  In contrast, the region near the Mn acceptor, labeled Mn, shows 
stronger $\sigma_{+}$ polarization than $\sigma_{-}$.  The energy dependence of the circular 
polarization $\rho_{\rm cir}$ is also shown.  It was computed from the two spectra using I$_{+}$(E) 
and I$_{-}$(E). The polarization changes sign at an energy of 1450~meV. This plot shows that the 
energy region dominated by A$^{\rm EM}$ recombination has \textit{negative} circular polarization, 
while the energy region dominated by A$^{\rm Mn}$ recombination has \textit{positive} polarization.

\begin{figure}
\begin{center}
\begin{tabular}{c}
\includegraphics[height=6.5cm]{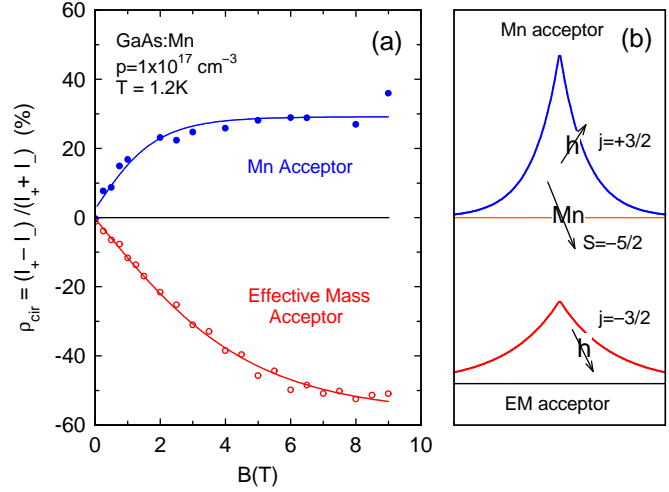}
\end{tabular}
\end{center}
\caption{(a) Circular polarization of the recombination photoluminescence from GaAs:Mn arising
from holes bound to effective mass acceptors and Mn acceptors.  (b) Diagram of the wavefunction of 
the hole in effective mass and Mn acceptors, showing the direction of hole and Mn spins in a
magnetic field pointing in the up direction.}
\end{figure}

Figure~5 shows the variation of the circular polarization as a function of magnetic field. For both
the A$^{\rm EM}$ and A$^{\rm Mn}$ recombination, the polarization increases linearly with
increasing magnetic field and shows saturation at high fields, due to a thermal activation process 
similar to a Brillouin function.  The reversal of the sign of the polarization indicates that the 
spin of the hole bound to a A$^{\rm Mn}$ is reversed relative to the hole bound to an A$^{\rm 
EM}$.\cite{Petrou1985} This is illustrated in the diagrams on the right of the figure. In the case 
of A$^{\rm EM}$, the valence band hole is spin down, j=--3/2. In the case of A$^{\rm Mn}$, the hole 
is spin up, j=+3/2. The exchange coupling between the Mn spin and the hole spin is responsible for 
reversing the direction of the hole. At low temperature, the MAC complex (hole+Mn) has the spin 
configuration J=1, m$_{\rm J}$= --1.  Thus, the exchange magnetic field from the oriented Mn spin 
overwhelms the applied magnetic field seen by the hole. There are other notable differences in the 
data for the two types of acceptors: the saturation value is smaller for A$^{\rm Mn}$, and the field 
required to reach saturation is smaller for A$^{\rm Mn}$. The exchange coupling between the hole 
spin and the Mn spin is discussed in the next section.

The temperature dependence of the circular polarization was investigated over the temperature
range from T=1.2 to 20~K. Figure~6a shows the circular polarization at temperatures from
T=1.56 to 20~K.  Higher temperatures require a larger magnetic field to produce saturation.
Figure~6b shows the dependence of the linear polarization at low temperature.  In contrast to the
circular polarization, there is no variation in the linear polarization above the noise level of a 
few degrees.  The small offset is related to the apparatus rather than the sample.  On the other 
hand, the A$^{\rm EM}$ recombination emission showed a small field dependence in the linear
polarization, but that is not of interest here.

\begin{figure}
\begin{center}
\begin{tabular}{c}
\includegraphics[height=9cm]{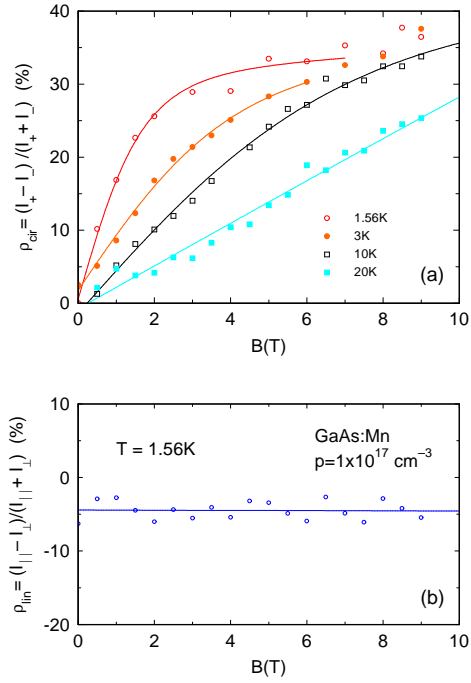}
\end{tabular}
\end{center}
\caption{Polarization of Mn acceptor recombination photoluminescence from GaAs:Mn versus
magnetic field for various temperatures.  Circular (a) and linear (b) polarizations are shown.}
\end{figure}

\section{MODELING}

\subsection{Polarization of Optical Transistions}
In order to fit the polarization data for the Mn acceptor it is necessary to determine the angular 
momentum mixing of the hole spin and Mn-ion spin, as well as optical transition matrix elements 
connecting the conduction band (\emph{cb}) and valence band states.  The angular momentum mixing can 
be determined using the Clebsch-Gordon coefficients (CG), while the matrix elements (ME) can be 
determined by considering the Luttinger-Kohn basis states.  These quantities are described in the 
following sections.  In addition to these two factors, overlap of the electron and hole 
wavefunctions ($\psi_{eh}$) could be important, but is neglected here because of the uncertainty in 
the wavefunction of the tightly bound hole.  The relative intensities of the emission in the various 
polarization configurations is thus proportional to
\[I \propto \left( {CG \times ME \times \psi_{eh} } \right)^2 .\]

In general, the magnetic field and temperature dependencies of the polarizations follow a 
Boltzman-like thermal activation function having an argument which is equal to the magnetic energy 
divided by the thermal energy.  At high fields and low temperatures (high B/T) the polarization 
saturates, as seen in Figs.~5~and~6.  The magnitude of the polarization at saturation is found
by assuming that the final electron states are $\psi_0|m_S,m_s\rangle=\psi_0|m_S\rangle|m_s\rangle$,
where $\psi_0$ is a \emph{n}=0 Landau level spatial wave function, $|m_S,m_s\rangle$ is a product
of a Mn ion (S=5/2) state with a conduction band (\emph{s}=1/2) basis state, and the initial hole
state is in the lowest-energy mixed acceptor state $|J,M_J\rangle=|1,+1\rangle$.\\

\subsubsection{Angular Momentum Mixing of Hole Spin and Mn ion Spin}
In the acceptor, mixing occurs between the j=3/2 hole spin and the S=5/2 Mn$^{2+}$ spin.
The total angular momentum of the hole-Mn complex is J=j+S=1,2,3,4.
Using Clebsch-Gordon coefficients, the eigenstates for the lowest energy state J=1 are:

\begin{center}
\[
\begin{array}{*{20}c}
   \hfill {\left| {J,M_J } \right\rangle } &\vline &  \hfill {\left| {m_S } \right\rangle \left| 
{m_j  = 3/2} \right\rangle } & \hfill {\left| {m_S } \right\rangle \left| {m_j  = 1/2} \right\rangle 
} & \hfill {\left| {m_S } \right\rangle \left| {m_j  =  - 1/2} \right\rangle } & \hfill {\left| {m_S 
} \right\rangle \left| {m_j  =  - 3/2} \right\rangle } \\
\hline
   \hfill {\left| {1, + 1} \right\rangle  = } &\vline &  \hfill { - \sqrt {1/20} \left| { - 1/2} 
\right\rangle \left| {3/2} \right\rangle } & \hfill {+\sqrt {3/20} \left| { + 1/2} \right\rangle 
\left| {1/2} \right\rangle } & \hfill { - \sqrt {3/10} \left| { + 3/2} \right\rangle \left| { - 1/2} 
\right\rangle } & \hfill {+\sqrt {1/2} \left| {5/2} \right\rangle \left| { - 3/2} \right\rangle } \\
   \hfill {\left| {1,\;\;\:0} \right\rangle  = } &\vline &  \hfill { - \sqrt {1/5} \left| { - 3/2} 
\right\rangle \left| {3/2} \right\rangle } & \hfill {+\sqrt {3/10} \left| { - 1/2} \right\rangle 
\left| {1/2} \right\rangle } & \hfill { - \sqrt {3/10} \left| { + 1/2} \right\rangle \left| { - 1/2} 
\right\rangle } & \hfill {+\sqrt {1/5} \left| {3/2} \right\rangle \left| { - 3/2} \right\rangle } \\
   \hfill {\left| {1, - 1} \right\rangle  = } &\vline &  \hfill { - \sqrt {1/2} \left| { - 5/2} 
\right\rangle \left| {3/2} \right\rangle } & \hfill {+\sqrt {3/10} \left| { - 3/2} \right\rangle 
\left| {1/2} \right\rangle } & \hfill { - \sqrt {3/20} \left| { - 1/2} \right\rangle \left| { - 1/2} 
\right\rangle } & \hfill {+\sqrt {1/20} \left| {1/2} \right\rangle \left| { - 3/2} \right\rangle } 
\\
\end{array}
\]
\end{center}

The final states in the \emph{cb} are products of Mn ion states $|m_{S}\rangle$ with Landau level
states for the \emph{cb}. The \emph{n}=0 Landau level state is $\psi_0|m_s\rangle$, where the
$m_s$=+1/2 state has the lower energy, since the effective g-factor, \emph{g}*=$-$0.44, is 
negative.\\

\subsubsection{Optical Transition Matrix Elements}
The interband optical transitions coupling electrons in the conduction band to \emph{free holes} in 
the valence band involve the standard matrix elements between Luttinger-Kohn basis states. The 
velocity operators, $v_\pm=v_x \pm v_y$, are in units of \emph{P}/\emph{m}, where \emph{P} is the
interband matrix element. These are given below for the two circular and linear polarizations,
$\sigma_\pm$ and $\sigma_\pi$, respectively.

\begin{center}
\[
\begin{array}{*{20}c}
   \hfill {Pol.} &\vline &  \hfill {\left\langle {s,m_s } \right|} & \hfill v \;\;\; & \hfill 
{\left| {j,m_j } \right\rangle }= & \hfill {ME} \\
\hline
   \hfill {\sigma _ +  } &\vline &  \hfill {\left\langle {1/2, + 1/2} \right|} & \hfill {v_ +  /2} & 
\hfill {\left| {3/2, - 1/2} \right\rangle  = } & \hfill {\sqrt {1/6} } \\
   \hfill {\pi}\;\:\: &\vline &  \hfill {\left\langle {1/2, + 1/2} \right|} & \hfill {v_z } \;\; & 
\hfill {\left| {3/2, + 1/2} \right\rangle  = } & \hfill {\sqrt {2/3} } \\
   \hfill {\sigma _ -  } &\vline &  \hfill {\left\langle {1/2, + 1/2} \right|} & \hfill {v_ -  /2} & 
\hfill {\left| {3/2, + 3/2} \right\rangle  = } & \hfill { - \sqrt {1/2} } \\
\hline
   \hfill {\sigma _ +  } &\vline &  \hfill {\left\langle {1/2, - 1/2} \right|} & \hfill {v_ +  /2} & 
\hfill {\left| {3/2, - 3/2} \right\rangle  = } & \hfill {\sqrt {1/2} } \\
   \hfill {\pi}\;\:\: &\vline &  \hfill {\left\langle {1/2, - 1/2} \right|} & \hfill {v_z } \;\; & 
\hfill {\left| {3/2, - 1/2} \right\rangle  = } & \hfill {\sqrt {2/3} } \\
   \hfill {\sigma _ -  } &\vline &  \hfill {\left\langle {1/2, - 1/2} \right|} & \hfill {v_ -  /2} & 
\hfill {\left| {3/2, + 1/2} \right\rangle  = } & \hfill { - \sqrt {1/6} } \\
\end{array}
\]
\end{center}

\subsubsection{Combined Angular Momentum Mixing and Optical Matrix Elements}
The following table shows the product of angular momentum mixing and optical matrix elements for 
transitions in which an electron in the n=0 \emph{cb} Landau level combines with a hole in the J=1 
magnetic acceptor complex (omitting a constant overlap integral, and \emph{P/m}).

\begin{center}
\[
\begin{array}{*{20}c}
   \hfill {\left\langle {m_S ,m_s } \right|} & \hfill v \;\; & \hfill {\left| {J,M_J } \right\rangle 
} &\vline &  \hfill {CG} \; & \hfill {ME} \;\;\;\; & \hfill {CG \times ME} \\
\hline
   \hfill {\left\langle {3/2,1/2} \right|} & \hfill {v_ +  /2} & \hfill {\left| {1,1} \right\rangle  
= } &\vline &  \hfill { - \sqrt {3/10} } & \hfill {\sqrt {1/6}  = } & \hfill { - \sqrt {1/20} } \\
   \hfill {\left\langle {1/2,1/2} \right|} & \hfill {v_z } \; & \hfill {\left| {1,1} \right\rangle  
= } &\vline &  \hfill {\sqrt {3/20} } & \hfill {\sqrt {2/3}  = } & \hfill {\sqrt {1/10} } \\
   \hfill {\left\langle { - 1/2,1/2} \right|} & \hfill {v_ -  /2} & \hfill {\left| {1,1} 
\right\rangle  = } &\vline &  \hfill { - \sqrt {1/20} } & \hfill { - \sqrt {1/2}  = } & \hfill 
{\sqrt {1/40} } \\
   \hfill {\left\langle {5/2, - 1/2} \right|} & \hfill {v_ +  /2} & \hfill {\left| {1,1} 
\right\rangle  = } &\vline &  \hfill {\sqrt {1/2} } & \hfill {\sqrt {1/2}  = } & \hfill {\sqrt {1/4} 
} \\
   \hfill {\left\langle {3/2, - 1/2} \right|} & \hfill {v_z } \; & \hfill {\left| {1,1} 
\right\rangle  = } &\vline &  \hfill { - \sqrt {3/10} } & \hfill {\sqrt {2/3}  = } & \hfill { - 
\sqrt {1/5} } \\
   \hfill {\left\langle {1/2, - 1/2} \right|} & \hfill {v_ -  /2} & \hfill {\left| {1,1} 
\right\rangle  = } &\vline &  \hfill {\sqrt {3/20} } & \hfill { - \sqrt {1/6}  = } & \hfill { - 
\sqrt {1/40} } \\
\hline
   \hfill {\left\langle {1/2,1/2} \right|} & \hfill {v_ +  /2} & \hfill {\left| {1,0} \right\rangle  
= } &\vline &  \hfill { - \sqrt {3/10} } & \hfill {\sqrt {1/6}  = } & \hfill { - \sqrt {1/20} } \\
   \hfill {\left\langle { - 1/2,1/2} \right|} & \hfill {v_z } \; & \hfill {\left| {1,0} 
\right\rangle  = } &\vline &  \hfill {\sqrt {3/10} } & \hfill {\sqrt {2/3}  = } & \hfill {\sqrt 
{1/5} } \\
   \hfill {\left\langle { - 3/2,1/2} \right|} & \hfill {v_ -  /2} & \hfill {\left| {1,0} 
\right\rangle  = } &\vline &  \hfill { - \sqrt {1/5} } & \hfill { - \sqrt {1/2}  = } & \hfill {\sqrt 
{1/10} } \\
   \hfill {\left\langle {3/2, - 1/2} \right|} & \hfill {v_ +  /2} & \hfill {\left| {1,0} 
\right\rangle  = } &\vline &  \hfill {\sqrt {1/5} } & \hfill {\sqrt {1/2}  = } & \hfill {\sqrt 
{1/10} } \\
   \hfill {\left\langle {1/2, - 1/2} \right|} & \hfill {v_z } \; & \hfill {\left| {1,0} 
\right\rangle  = } &\vline &  \hfill { - \sqrt {3/10} } & \hfill {\sqrt {2/3}  = } & \hfill { - 
\sqrt {1/5} } \\
   \hfill {\left\langle { - 1/2, - 1/2} \right|} & \hfill {v_ -  /2} & \hfill {\left| {1,0} 
\right\rangle  = } &\vline &  \hfill {\sqrt {3/10} } & \hfill { - \sqrt {1/6}  = } & \hfill { - 
\sqrt {1/20} } \\
\hline
   \hfill {\left\langle { - 1/2,1/2} \right|} & \hfill {v_ +  /2} & \hfill {\left| {1, - 1} 
\right\rangle  = } &\vline &  \hfill { - \sqrt {3/20} } & \hfill {\sqrt {1/6}  = } & \hfill { - 
\sqrt {1/40} } \\
   \hfill {\left\langle { - 3/2,1/2} \right|} & \hfill {v_z } \; & \hfill {\left| {1, - 1} 
\right\rangle  = } &\vline &  \hfill {\sqrt {3/10} } & \hfill {\sqrt {2/3}  = } & \hfill {\sqrt 
{1/5} } \\
   \hfill {\left\langle { - 5/2,1/2} \right|} & \hfill {v_ -  /2} & \hfill {\left| {1, - 1} 
\right\rangle  = } &\vline &  \hfill { - \sqrt {1/2} } & \hfill { - \sqrt {1/2}  = } & \hfill {\sqrt 
{1/4} } \\
   \hfill {\left\langle {1/2, - 1/2} \right|} & \hfill {v_ +  /2} & \hfill {\left| {1, - 1} 
\right\rangle  = } &\vline &  \hfill {\sqrt {1/20} } & \hfill {\sqrt {1/2}  = } & \hfill {\sqrt 
{1/40} } \\
   \hfill {\left\langle { - 1/2, - 1/2} \right|} & \hfill {v_z } \; & \hfill {\left| {1, - 1} 
\right\rangle  = } &\vline &  \hfill { - \sqrt {3/20} } & \hfill {\sqrt {2/3}  = } & \hfill { - 
\sqrt {1/10} } \\
   \hfill {\left\langle { - 3/2, - 1/2} \right|} & \hfill {v_ -  /2} & \hfill {\left| {1, - 1} 
\right\rangle  = } &\vline &  \hfill {\sqrt {3/10} } & \hfill { - \sqrt {1/6}  = } & \hfill { - 
\sqrt {1/20} } \\
\end{array}
\]
\end{center}

\bigskip
\subsubsection{Wavefunction Overlap}
In addition to the angular momentum mixing and optical matrix elements, there is also a factor 
involving the overlap of the hole wavefunction $\psi(\textbf{r})$=$f(\textbf{r})$ with the Landau 
wavefunctions of the conduction band. Several theories of acceptors incorporating central cell 
attraction have been developed.\cite{Bhattacharjee1999,Malyshev1997}  Baldereschi and Lipari have 
modeled the acceptor without central cell attraction.\cite{Baldereschi1974} In this model, the 
\emph{acceptor-bound hole} wavefunctions in the spherical approximation are given 
by:\cite{Baldereschi1973}
\begin{center}
\[
\begin{array}{*{20}c}
   {\left| {j,m_j } \right\rangle } &\vline &  {m_j  = 3/2} & {m_j  = 1/2} & {m_j  =  - 1/2} & {m_j  
=  - 3/2}  \\
\hline
   {\left| {3/2, + 3/2} \right\rangle  = } &\vline &  {[f \; Y_0^0  + \sqrt {1/5} gY_2^0 } & { - 
\sqrt {2/5} g \; Y_2^1 } & {\sqrt {2/5} g \; Y_2^2 } & {0]}  \\
   {\left| {3/2, + 1/2} \right\rangle  = } &\vline &  {[\sqrt {2/5} g \; Y_2^{ - 1} } & {f \; Y_0^0  
- \sqrt {1/5} gY_2^{ 0} } & 0 & {\sqrt {2/5} g \; Y_2^2 ]}  \\
   {\left| {3/2, - 1/2} \right\rangle  = } &\vline &  {[\sqrt {2/5} g \; Y_2^{ - 2} } & 0 & {f \; 
Y_0^0  - \sqrt {1/5} gY_2^0 } & {\sqrt {2/5} g \; Y_2^1 ]}  \\
   {\left| {3/2, - 3/2} \right\rangle  = } &\vline &  {[0} & {\sqrt {2/5} g \; Y_2^{ - 2} } & { - 
\sqrt {2/5} g \; Y_2^{ - 1} } & {f \; Y_0^0  + \sqrt {1/5} gY_2^0 ]}  \\
\end{array}
,\]
\end{center}
where the multicomponent wavefunctions on the right side are usually represented as column vectors 
instead of row vectors.

\bigskip
\bigskip
\section{Results of Polarization Saturation}
From Fig.~6 we find the experimental circular and linear polarizations at saturation, listed in the 
table below.  The theoretical values were computed using the coefficients in the first three rows of 
the table in section 4.1.3.
\[\begin{array}{*{20}c}
   {Polarization} &\vline &  {} &\vline &  {Experiment} &\vline &  {Theory}  \\
\hline
\vspace {0.1cm}
   {\rho _{circ} } &\vline & { = {{I_+  - I_- } \over {I_+  + I_- }}} &\vline &  {0.35 \pm 0.05} 
&\vline &  {0.33}  \\
\hline
\vspace {0.1cm}
   {\rho _{lin} } &\vline &  { = {{I_\parallel  - I_+   - I_-  } \over {I_\parallel   + I_+   + I_-  
}}} &\vline &  {0.04 \pm 0.05} &\vline &  {0.14}  \\
\hline
\end{array}\]
For circular polarization there is good agreement between experiment and theory.  For linear 
polarization the agreement is not as good.  However, by assuming a smaller wavefunction overlap for 
$\pi$-polarization, where $\psi_{eh}^2(\pi)$=$\frac{3}{4}\psi_{eh}^2(\sigma)$, then $\rho_{lin}$=0 
is in better agreement with the experiment.


\bigskip
\section{SUMMARY}
The emission arising from recombination of electrons with holes bound in Mn acceptors in 
lightly-doped GaAs:Mn becomes polarized in an applied magnetic field.  In a Faraday configuration, 
the circular polarization saturates at high fields at a value of $\rho_{circ}$=1/3.  On the other 
hand, in a Voigt configuration, the linear polarization has near-zero polarization.  These values 
are in agreement with theory by taking into account: (\emph{i}) the angular momentum mixing of the 
3/2 hole-spin with the 5/2 Mn spin; and (\emph{ii}) the optical transition matrix elements 
connecting the conduction and valence band states.  It was found that there is little effect on the 
polarization due to overlap of the electron and hole wavefunctions.\\

\section*{ACKNOWLEDGMENTS}
This work was supported by the NSF grant DMR-0305360.  Samples were provided by J. Hao and K.C.
Hsieh.\\


\bibliography{heiman}
\bibliographystyle{spiebib}   

\end{document}